\documentclass[journal,twoside,web]{ieeecolor}
\usepackage{tmi}
\usepackage{cite}
\usepackage{amsmath,amssymb,amsfonts}
\usepackage{graphicx}
\usepackage{textcomp}
\usepackage{algorithm}
\usepackage{algpseudocode}
\usepackage{soul}
\usepackage{xcolor}
\usepackage{setspace}

\usepackage[switch]{lineno}
\usepackage{amsmath}
\usepackage{comment}

\def\BibTeX{{\rm B\kern-.05em{\sc i\kern-.025em b}\kern-.08em
    T\kern-.1667em\lower.7ex\hbox{E}\kern-.125emX}}
\begin{document}
\title{Morphology-based non-rigid registration of coronary computed tomography and intravascular images through virtual catheter path optimization}
\author{Karim Kadry, Max L. Olender, Andreas Schuh, Abhishek Karmakar, Kersten Petersen, Michiel Schaap, David Marlevi, Adam UpdePac, Takuya Mizukami, Charles Taylor, Elazer R. Edelman, and Farhad R. Nezami}

\maketitle

\begin{abstract}
Coronary computed tomography angiography (CCTA) provides 3D information on obstructive coronary artery disease, but cannot fully visualize high-resolution features within the vessel wall. Intravascular imaging, in contrast, can spatially resolve atherosclerotic in cross sectional slices, but is limited in capturing 3D relationships between each slice. Co-registering CCTA and intravascular images enables a variety of clinical research applications but is time consuming and user-dependent. This is due to intravascular images suffering from non-rigid distortions arising from irregularities in the imaging catheter path. To address these issues, we present a morphology-based framework for the rigid and non-rigid matching of intravascular images to CCTA images. To do this, we find the optimal virtual catheter path that samples the coronary artery in CCTA image space  to recapitulate the coronary artery morphology observed in the intravascular image. We validate our framework on a multi-center cohort of 40 patients using bifurcation landmarks as ground truth for longitudinal and rotational registration. Our registration approach significantly outperforms other approaches for bifurcation alignment. By providing a differentiable framework for  multi-modal vascular co-registration, our framework reduces the manual effort required to conduct large-scale multi-modal clinical studies and enables the development of machine learning-based co-registration approaches.
\end{abstract}
\thanks{ This work was partially supported by the National Institute of Health (1R01HL161069) to ERE and FRN and Heartflow, Inc.

KK, MLO, ERE are with the Institute of Medical Engineering, Massachussetts Institute of Technology, Cambridge, MA, USA (emails: kkadry@mit.edu, molender@mit.edu, ere@mit.edu)

AS, KP, MS, AU, and CT are with HeartFlow, Inc., Redwood City, CA, 94063,  USA (emails: aschuh@heartflow.com, kpetersen@heartflow.com, mschaap@heartflow.com, aupdepac@heartflow.com, ctaylor@heartflow.com)

AK is with the Meinig School of Biomedical Engineering, Cornell University, Ithaca, NY 14850 (email: ak944@cornell.edu)

DM is with the Department of Molecular Medicine and Surgery, Karolinska Institute, Stockholm, Sweden,(email: david.marlevi@ki.se)

TM is with the Cardiovascular Center in Aalst, OLV Clinic, Aalst, Belgium, (email: takuyamizukami@coreaalst.com)

FRN is with the Department of Surgery Brigham and Women's Hospital Harvard Medical School Boston, MA 02115, (email: frikhtegarnezami@bwh.harvard.edu)
}

\begin{IEEEkeywords}
Slice-to-volume registration, Image registration, free-form deformation, spatial transforms, optical coherence tomography, multi-modal data fusion.
\end{IEEEkeywords}
\section{Introduction}
\label{sec:introduction}
Coronary computed tomography angiography (CCTA) is a 3D imaging modality that allows for the detection of stenotic atherosclerotic lesions and assists clinicians in the diagnosis and treatment of coronary artery disease (CAD). {\color{black} In contrast to the current gold standard of digital subtraction angiography (DSA)}, CCTA can be used to create 3D computational models of coronary blood flow that can estimate fractional flow reserve (FFR-CT), \cite{uzu2019lumen}. CCTA also provides information on soft-tissue intraplaque components within the wall, albeit with some limitations. For example, CCTA suffers from blooming artifacts in the presence of highly attenuating calcium deposits \cite{kim2015limitations,budoff2008diagnostic}, which, combined with comparably low image resolution, creates difficulties in resolving highly calcified arteries.  In contrast, catheter based imaging modalities such as intravascular ultrasound (IVUS) and optical coherence tomography (OCT), provide high-fidelity cross-sectional images of the lumen and intra-plaque. {\color{black}  However, catheter based modalities do not contain information on the 3D pose (location and orientation) for each frame, making it difficult to reconstruct the artery in 3D.
Recovering the pose of each intravascular frame within the CCTA image is known as co-registration, and enables three key clinical applications. First, intravascular image slices can be directly used as ground-truth in clinical studies to study the viability of CCTA in assessing CAD-related diagnostic metrics such as luminal area \cite{uzu2019lumen}, calcium morphology \cite{takahashi2021feasibility}, and plaque burden \cite{fischer2013coronary,de2013automaticivusCT,brodoefel2009accuracyCTIVUS}. Second, co-registration enables the creation of matched multi-modal datasets, which can be used to train neural networks for the segmentation of lumen and plaque within CCTA images. Third, high-fidelity segmentations derived from intravascular images can be used in tandem with the recovered poses to create high fidelity coronary digital twins \cite{kadry2021platform,straughan2023fully,van2019automatedelucid}. Such patient-specific models enable the physics based simulation of various biophysical phenomena such as hemodynamics \cite{uzu2019lumen}, biomechanical pressurization \cite{kadry2021platform,straughan2023fully}, and virtual interventions \cite{poletti2022towards,rikhtegar2016drug}, which guides clinical decision making and pathophysiological research.}

Manual co-registeration of CCTA and intravascular images is however, a challenging and time consuming task. Typically, cross-sectional frames of interest are extracted from the CCTA images which then have to be matched with corresponding frames from a catheter-based intravascular acquisition\cite{takahashi2021feasibility,fischer2013coronary,uzu2019lumen,brodoefel2009accuracyCTIVUS}. Rigid registration in the longitudinal and rotational directions is usually achieved by matching single landmarks in both modalities, such as large bifurcations \cite{takahashi2021feasibility}. However, the beating of the heart, the irregular motion of the imaging catheter, and the rotation of the catheter about its own axis create non-rigid distortions that accumulate along the length of the pullback \cite{tsiknakis2021ivus}. Manually correcting for such artifacts is prohibitively time-consuming, requiring a cardiologist to mark multiple fiduciary points in both images and locate the equivalent frames accordingly. There is therefore a need for computational algorithms that non-rigidly register CCTA images to corresponding intravascular data in an automatic fashion.
%Manually correcting for such artifacts is prohibitively time-consuming, requiring a cardiologist to mark multiple fiduciary points in both images and shift one modality such that the annotated points sufficiently align (\cite{carlier2014new,tu2011fusion,hebsgaard2015co})

%\section{Related Work}
%\subsection{Rigid Registration of intravascular images}
Automatic co-registration techniques for longitudinal alignment typically consist of discretely optimizing a cost function over a set of longitudinal or rotational image shifts, where the cost function varies depending on the modalities being registered. Some proposed cost functions include metrics such as lumen diameters \cite{qin2021automatic}, lumen contours \cite{molony2016evaluation,karmakar2020detailed}, calcium thickness \cite{gharaibeh2020co,molony2016evaluation}, and image pixel intensities \cite{tsiknakis2021ivus}. {\color{black} In addition to longitudinal co-registration, our prior work includes} rigid rotational registration for intravascular pullbacks based on extracted features such as luminal contours \cite{karmakar2020detailed}. However, the registration accuracy of all rigid registration methods is compromised by inconsistent motor pullback speeds, rotational drift, and cardiac motion, as these introduce non-rigid longitudinal and rotational distortions that misalign image features such as diseased plaque and bifurcations.

%\subsection{Non-rigid Registration}
To compensate for the longitudinal, rotational, and transverse motion of the catheter, several non-rigid registration approaches have been proposed. Non-rigid registration of intravascular imaging datasets has been predominantly performed through dynamic time warping (DTW) and dynamic programming (DP) \cite{tsiknakis2021ivus,molony2016evaluation,karmakar2022framework}. However, DTW introduces non-physiological assumptions into the registration process by discretely skipping or repeating intravascular frames, assumed to be evenly spaced along the longitudinal direction.{\color{black} In contrast to discrete approaches, previous works, including our own}, have leveraged continuous non-rigid registration methods to model the longitudinal stretch and rotational drift between intravascular imaging frames using affine transforms and spline interpolation \cite{zhang2014side,uzu2019lumen}. While such continuous non-rigid methods are more realistic, they extensively rely on manual pre-processing and the annotation of all bifurcation zones for image registration  {\color{black} and do not account for the bending of the catheter away from the vessel centerline.}

Further, there has been an increasing interest in machine learning approaches to image co-registration in which a neural network is trained to predict a spatial transform that maps a moving image onto a static target image \cite{balakrishnan2019voxelmorph,fu2020deepreview,gopalakrishnan2023intraoperative}. Such approaches critically rely on  differentiable spatial transforms and rendering operations for the back-propagation of gradients to adjust the neural network weights \cite{jaderberg2015spatialtrans,gopalakrishnan2022fast}. While such transforms are available for co-registration of 3D medical images in rectangular coordinates \cite{balakrishnan2019voxelmorph}, a similar  framework that accounts for the unique variation in intravascular catheter motion has yet to be developed.

 {\color{black}Given the previous limitations in prior approaches, we here propose a novel slices-to-volume registration framework that aligns a set of intravascular image slices to their equivalent location in a volumetric CCTA image}. The proposed continuous registration methodology does not require manual matching of morphological landmarks, requires only the morphology (lumen and vessel wall) for both modalities, along with the centerline within the CCTA image space. Specifically, we explore the problem of reconstructing the path of a \emph{virtual catheter} sampling from a 3D CCTA-derived lumen morphology such that the cross sectional slices sampled by the virtual catheter match the image slices from the equivalent intravascular pullback. We specifically demonstrate our algorithm in the case of OCT intravascular pullbacks, where our key contributions are as follows:

\begin{itemize}
  %\item A continuous co-registration framework for rigid and non-rigid matching of CCTA images and intravascular images up to pixelwise alignment, with segmentations of the lumen and vessel wall and CT centerline as sole input.
%  \item We leverage a rigid initialization approach that consists of our published longitudinal rigid registration algorithm, which uses lumen area in a multi-step decision process, and a rotational registration step that leverages the segmentation of the vessel wall to produce an initial rotational configuration for subsequent refinement.
  \item  {\color{black} We introduce a differentiable and non-rigid spatial transform that acts on a set of frames defining the path of a virtual catheter in 3D space. The transform is formulated in terms of intravascular catheter motion, specifically modelling longitudinal, rotational, and transverse distortions. Our spatial transform is regularized to enforce priors regarding cumulative motion distortions and smoothness, while also being compatible with deep learning registration frameworks. 
  \item We propose a rigid and non-rigid registration procedure for intravascular image slices and CCTA volumes, based on matching lumen and vessel wall morphology between modalities. The virtual catheter is initialized by the rigid registration step and then stretched, twisted, and bent by the non-rigid step through gradient-based optimization. For non-rigid registration, we choose to optimize the similarity between luminal distance fields and introduce several pre-processing steps to stabilize the process.
  \item We demonstrate the capabilities of our registration procedure on a multi-center dataset of 40 CCTA and OCT images with manually annotated landmarks. We directly benchmark against our previously developed discrete optimization approach and demonstrate improved registration error.}

\end{itemize}

\section{Methodology}
\begin{figure*}[!ht]
\centerline{\includegraphics[width=0.9\textwidth]{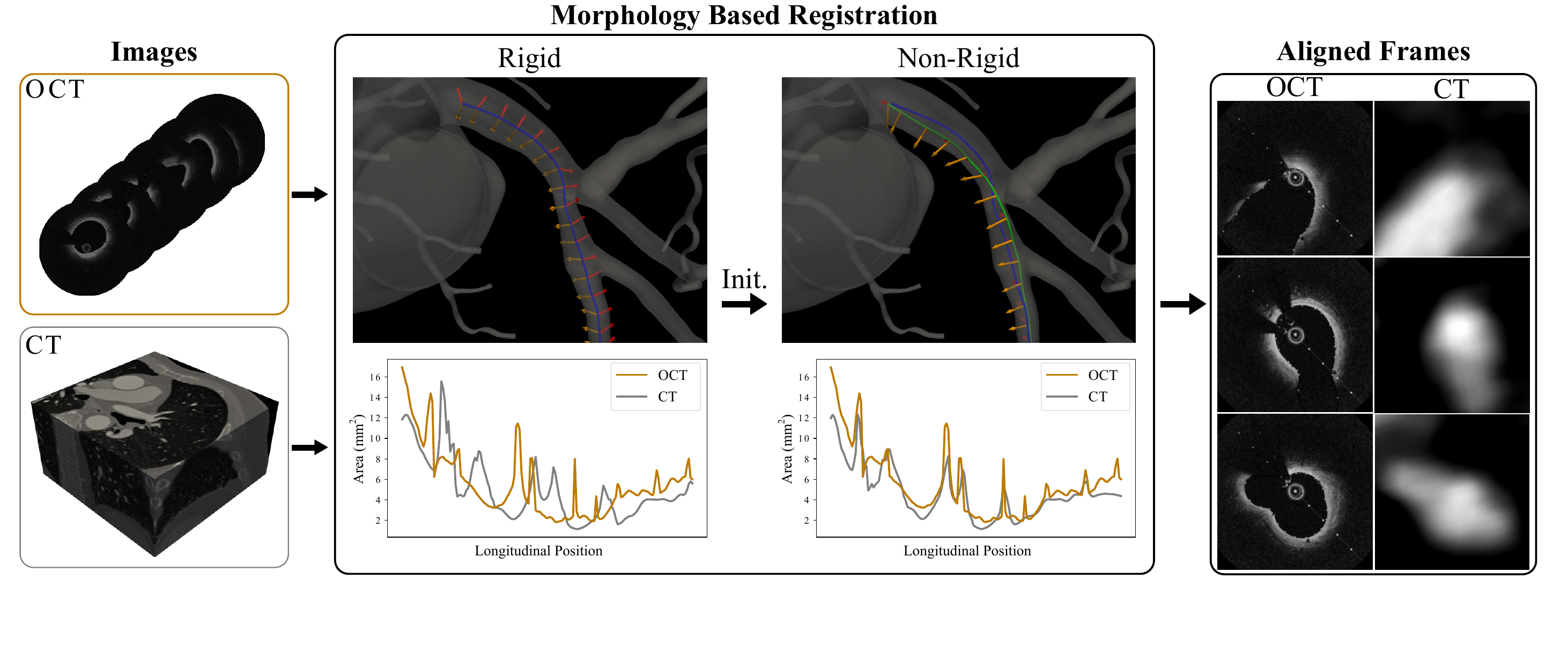}}
\caption{Overview of the proposed registration pipeline. Lumen and vessel wall morphologies are  {\color{black} derived from OCT and CCTA images and given as input to the registration process along with an initial centerline centerline in CCTA space. A  rigid co-registration step initializes the virtual catheter path (blue), where the pose of each frame is described by orientation vectors (red and orange arrows). The virtual catheter path is used to sample a virtual pullback of the CCTA morphology in the form of a luminal distance field. The morphological similarity between the OCT and CCTA pullbacks is used to guide the non-rigid registration step. The alignment process stretches, twists, and bends the virtual catheter frames to produce an aligned catheter path (green) that can sample equivalent CCTA frames for each intravascular image slice.}}
\label{figurenonrigid}
\end{figure*}
%\subsection{Overview of coregistration framework}
An overview of the co-registration pipeline is detailed in Fig. \ref{figurenonrigid}. In brief, our registration algorithm \ref{sec:coreg_overview} takes as input morphological representations of lumen and vessel wall for the CCTA and intravascular images, in addition to the CCTA lumen centerline. In this study, we utilize OCT pullbacks as our intravascular imaging modality. For rigid registration, a virtual catheter is first initialized from the centerline in the form of 3D frame positions and poses detailing the orientation of each frame. These frames are used to sample cross sections from the CCTA morphology to produce a virtual pullback (section \ref{VirtualCatheter}). The sampled pullback is used for longitudinal and rotational alignment (section \ref{Rigid Registration}), which outputs crop indices and a rotation angle that initialize the non-rigid registration process. Non-rigid registration (section \ref{nonrigid_methods_overview}) optimizes a spatial transform applied to the virtual catheter that aligns the morphology between the virtual and intravascular pullbacks. The non-rigid spatial transformation consists of longitudinal (section \ref{nonrigid_methods_long}), rotational (section \ref{nonrigid_methods_rot}), and transverse (section \ref{nonrigid_methods_trans}) deformation steps. To evaluate our method, we usemorphological representations derived from  a multi-center clinical image dataset and evaluate the performance of our algorithm against discrete optimization baselines (section \ref{sec:evaluation}).

% In brief, our algorithm takes requires 5 inputs 1) the CCTA centerline frames, 2) the CCTA lumen morphology, 3) the CCTA vessel wall morphology, 4) the OCT lumen morphology, and 5) the OCT vessel wall morphology. 

% \newcommand\Algphase[1]{%
% \vspace*{-.7\baselineskip}\Statex\hspace*{\dimexpr-\algorithmicindent-2pt\relax}\rule{\columnwidth}{0.4pt}%
% \Statex\hspace*{-\algorithmicindent}\textbf{#1}%
% \vspace*{-.7\baselineskip}\Statex\hspace*{\dimexpr-\algorithmicindent-2pt\relax}\rule{\columnwidth}{0.4pt}%
% }

\newcommand\Algphase[1]{
\vspace*{-0.5\baselineskip}
\Statex\hspace*{\dimexpr-\algorithmicindent-2pt\relax}
\rule{\columnwidth}{0.4pt}%
\vspace*{-0.2\baselineskip}
\Statex\hspace*{-\algorithmicindent}
\textbf{#1}%
\vspace*{-.7\baselineskip}
\Statex\hspace*{\dimexpr-\algorithmicindent-2pt\relax}
\rule{\columnwidth}{0.4pt}%
}

\subsection{Co-registration framework}
\label{sec:coreg_overview}
\subsubsection{Input morphological representations}
\label{Preprocess}

 {\color{black}  As CCTA and intravascular images are dissimilar in their image characteristics, we choose to align the images based on morphological representations of the lumen and vessel wall. The four input morphologies consist of the luminal Signed Distance Fields (SDFs) ($\mathbf{L}_\text{CT}^\text{3D}, \mathbf{L}_\text{OCT}^\text{pull}$), as well as the vessel wall SDFs ($\mathbf{W}_\text{CT}^\text{3D}, \mathbf{W}_\text{OCT}^\text{pull}$) for both modalities. The superscripts '3D' and 'pull' indicate whether the SDF is located in 3D cartesian space or the cylindrical space defined by the catheter respectively.}

\subsubsection{Virtual catheterization}
\label{VirtualCatheter}
 {\color{black}To compare both modalities in the same coordinate system, we leverage curved-planar reformation \cite{kanitsar2002cpr}, where a virtual catheter samples cross-sectional slices from the CCTA lumen and vessel wall to produce $\mathbf{L}_\text{CT}^\text{pull}$ and $\mathbf{W}_\text{CT}^\text{pull}$ respectively. The virtual catheter is defined by a set of frames in 3D space that are constructed through a two step process that takes as input the set of $n$ CCTA centerline points $\mathbf{R}\in\mathbb{R}^{n\times3}$ arranged in 3D space. The first step consists of finding the set of tangent vectors $\textbf{T}\in\mathbb{R}^{n\times3}$ by applying a spatial derivative on $\mathbf{R}$, defining normal vectors for each frame. The second step consists of finding the orthogonal orientation vectors $\textbf{U}\in\mathbb{R}^{n\times3}$ and $\textbf{V}\in\mathbb{R}^{n\times3}$ that define the angular orientation of each frame. This is done through randomly initializing the orthogonal vectors for the first frame and applying parallel transport \cite{guo2013meshpartrans} along the centerline, which ensures that all orientation vectors remain stable between frames. The frames $\mathbf{F}\in\mathbb{R}^{n\times3\times4}$ are finally obtained by concatenating the position and orientation vectors,

\begin{equation}
\mathbf{F}=\text{concat}(\mathbf{R},\mathbf{T},\mathbf{U},\mathbf{V}).
\end{equation} To produce a virtual pullback, the frame matrix $\mathbf{F}$ is represented as a set of planar pointclouds and is used to sample the CCTA SDFs ($\mathbf{L}_\text{CT}^\text{3D}$,$\mathbf{W}_\text{CT}^\text{3D}$) for each point, where the cross-sectional size of the resulting grid matches that of the intravascular dataset,
\begin{equation}
    \mathbf{L}_\text{CT}^\text{pull}=\text{VirtualCatheter}(\mathbf{F},\mathbf{L}_\text{CT}^\text{3D}),
\end{equation}
\begin{equation}
    \mathbf{W}_\text{CT}^\text{pull}=\text{VirtualCatheter}(\mathbf{F},\mathbf{W}_\text{CT}^\text{3D}).
\end{equation}}

\subsubsection{Rigid registration}

\label{Rigid Registration}

\begin{algorithm}
\caption{Full Co-registration Algorithm}
\label{alg:full_coreg}
\begin{algorithmic}[1]\setstretch{1.3}
% \vspace{-1mm}
 \color{black}
\small
\Require $\mathbf{R}^\text{ori}$ \Comment{CCTA Original Centerline Points}

\Require $\mathbf{L}_{\text{CT}}^\text{3D},
\mathbf{L}_{\text{OCT}}^\text{pull}$ \Comment{Luminal Signed Distance Fields}

\Require $\mathbf{W}_{\text{CT}}^\text{3D},
\mathbf{W}_{\text{OCT}}^\text{pull}$ \Comment{Wall Signed Distance Fields}

\Algphase{Initialize Pullback from CCTA Morphology (Sec. \ref{VirtualCatheter})}

\State $\mathbf{F}^\text{ori} \leftarrow \text{InitFrames}(\mathbf{R}^\text{ori})$ \Comment{Frame Positions \& Poses}

\State $\mathbf{L}_{\text{CT}}^\text{pull} \leftarrow \text{VirtualCatheter}(\mathbf{F}^\text{ori}, \mathbf{L}_{\text{CT}}^{\text{3D}})$ 

\State $\mathbf{W}_{\text{CT}}^\text{pull} \leftarrow \text{VirtualCatheter}(\mathbf{F}^\text{ori}, \mathbf{W}_{\text{CT}}^{\text{3D}})$

\Algphase{Rigid Registration w/ Lumen \& Vessel Wall (Sec. \ref{Rigid Registration})}

% \State $\mathbf{F}, \mathbf{L}_{\text{OCT}}^{\text{pull}},  \mathbf{W}_{\text{OCT}}^{\text{pull}}\leftarrow$
\State $\mathbf{C} \leftarrow \text{LongReg}(\mathbf{L}_{\text{CT}}^{\text{pull}}, \mathbf{L}_{\text{OCT}}^{\text{pull}})$  \Comment{Crop Indices}

\State $\vartheta \leftarrow \text{RotReg}(\mathbf{C}, \mathbf{W}_{\text{CT}}^{\text{pull}}, \mathbf{W}_{\text{OCT}}^{\text{pull}})$  \Comment{Rotation Angle}

% \State $\mathbf{F}\leftarrow \text{Rotate(}\mathbf{F})$
\Algphase{Non-rigid Registration w/ Lumen (Sec. \ref{nonrigid_methods_overview})}
\State $\mathbf{F}_{\varphi} \leftarrow \text{NonrigidReg}(\mathbf{C}, \vartheta,\mathbf{F}^\text{ori}, \mathbf{L}_{\text{CT}}^\text{3D},\mathbf{L}_{\text{OCT}}^{\text{pull}})$

\State \Return $F_{\varphi} $
\end{algorithmic}
\end{algorithm}

\begin{figure*}[!h]
\centerline{\includegraphics[width=0.8\textwidth]{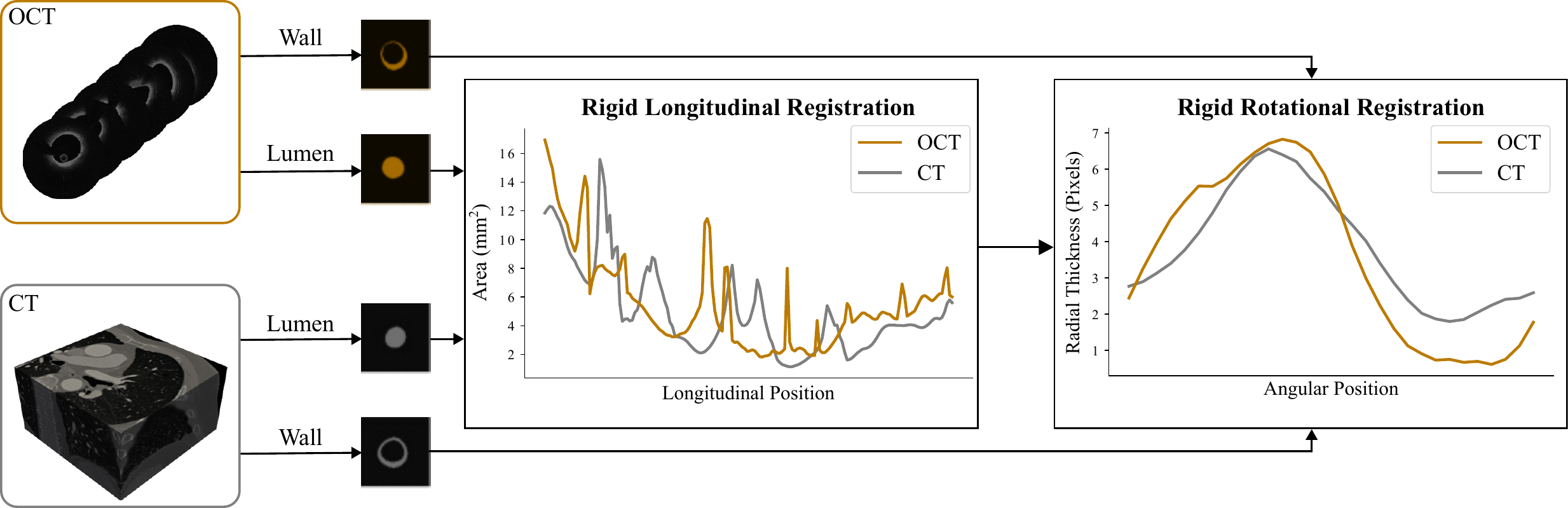}}
\caption{Overview of the proposed rigid registration pipeline. Lumen area vectors from both modalities are used for rigid registration in the longitudinal direction using a sliding window approach. The longitudinal registration is used to crop each segmentation such that they have the same starting point for rotational registration. The vessel wall segmentations for all frames are converted to vessel thickness-angle plots and are used to determine a single optimal rotation for the entire pullback.}
\label{figurerigid}
\end{figure*}
An overview of the rigid registration step can be seen in Fig. \ref{figurerigid} and Algorithm \ref{alg:full_coreg}.  {\color{black} We first construct the virtual catheter frames $\mathbf{F}^\text{ori}$ from the input centerline points $\mathbf{R}_\text{ori}$ and use them to initialize the virtual pullbacks $\mathbf{L}_\text{CT}^\text{pull}$ and $\mathbf{W}_\text{CT}^\text{pull}$, which are used for longitudinal and rotational registration respectively. For the rigid longitudinal registration, we binarize the luminal SDFs ($\mathbf{L}_\text{CT}^\text{pull}$ \& $\mathbf{L}_\text{OCT}^\text{pull}$) }and create area vectors for each modality. We leverage our previous work to rigidly align the pullbacks using a multi-step sliding window method, minimizing the difference in area vectors and bifurcation locations \cite{karmakar2020detailed}. The resulting output consists of cropping indices $\mathbf{C}=\{\mathbf{C}_{CT},\mathbf{C}_{OCT}\}$ determining the shared starting points for each modality.
\begin{equation}
\color{black}
    \mathbf{C}=\text{LongReg}(\mathbf{L}_{\text{CT}}^{\text{pull}}, \mathbf{L}_{\text{OCT}}^{\text{pull}}),
\end{equation}

For rigid rotational registration, using luminal profiles for rigid rotational alignment was deemed unreliable due to the CCTA-derived morphology being circularly symmetric.  {\color{black} Therefore, we take as input the vessel wall SDFs ($\mathbf{W}_\text{CT}^\text{pull}$, $\mathbf{W}_\text{OCT}^\text{pull}$) that were correspondingly binarized to produce segmentation maps. For both wall segmentations, a wall thickness matrix $\mathbf{H} \in \mathbb{R}^{n\times \gamma}$ by tracing $\gamma$ radial rays from the centroid of all $n$ frames of the vessel segmentation in equally spaced circumferential increments. We crop the thickness matrices for the CCTA ($\mathbf{H}_\text{CT}$) and OCT ($\mathbf{H}_\text{OCT}$) images using the indices $\mathbf{C}$ such that they are longitudinally aligned with the same starting points. The optimal rigid rotation angle $\vartheta$ is obtained by circumferentially sliding one thickness matrix over the other and minimizing the mean squared error,
\begin{equation}
    \vartheta=\text{RotReg}(\mathbf{W}_{\text{CT}}^{\text{pull}}, \mathbf{W}_{\text{OCT}}^{\text{pull}}).
\end{equation}}

\subsubsection{Non-rigid registration overview}
\label{nonrigid_methods_overview}

The non-rigid registration process can be seen in Fig. \ref{figur_Virtual} and Algorithm \ref{alg:nonrig_coreg}.  {\color{black} The input consists of the initialized frame variables $\mathbf{F}^\text{ori}$, cropping indices $\mathbf{C}$, rotation angle $\vartheta$, and the luminal SDFs ($\mathbf{L}_\text{CT}^\text{3D}$, $\mathbf{L}_\text{CT}^\text{pull}$). We formulate the problem in terms of finding the set of frames $\mathbf{F}^\varphi$ that correspond to the original path of the intravascular catheter in CCTA image space. This is done by maximizing the morphological similarity between the intravascular pullback and the CCTA virtual pullback sampled with spatially transformed frames $\mathbf{F}^\varphi$. First, the rigidly initialized frames $\mathbf{F}$ are obtained by cropping and rotating the input frames $\mathbf{F}^\text{ori}$ according to the outputs of rigid registration ($\mathbf{C}$,$\vartheta$).} The updated frame variables $\textbf{F}^\varphi$ are produced through three sequentially applied non-rigid spatial transforms $\varphi_\text{long}$, $\varphi_\text{rot}$, and $\varphi_\text{trans}$ in the longitudinal, rotational, and transverse, directions respectively: 
\begin{equation}
    \textbf{F}^\varphi=\varphi_\text{trans}\circ\varphi_\text{rot}\circ \varphi_\text{long}\circ\textbf{F},
\end{equation}
{\color{black} where $\circ$ is the composition operator. The morphological similarity function was defined as the mean squared error between the CCTA morphology $\mathbf{L}_\text{CT}^\text{pull}$ sampled with the spatially transformed frames $\textbf{F}^\varphi$ and the OCT morphology $\mathbf{L}_\text{OCT}^\text{pull}$ that is considered as a target}. We specifically clamp the SDFs to only have non-zero values inside the lumen to prevent the virtual catheter from switching to the incorrect coronary branch:
\begin{equation}
\color{black}
    \mathcal{L}=MSE\bigl(\text{clamp}(\mathbf{L}_{\text{CT}}^{\text{pull}}),\text{clamp}( \mathbf{L}_{\text{OCT}}^{\text{pull}})\bigl).
\end{equation}
This approach was used instead of a segmentation-based similarity function, such as cross-entropy or Dice, as binary-segmentation-based losses reach a minimum value when there is complete overlap between segmentations and thus are poor surrogates for alignment \cite{rohlfing2011image_similarity}. In contrast, distance field-based losses continue to change even after complete overlap is achieved, allowing for enhanced registration accuracy.

\subsubsection{Virtual Catheter Manipulation}
\label{sec:manipulation}
\begin{figure*}[!h]
\centerline{\includegraphics[width=0.9\textwidth]{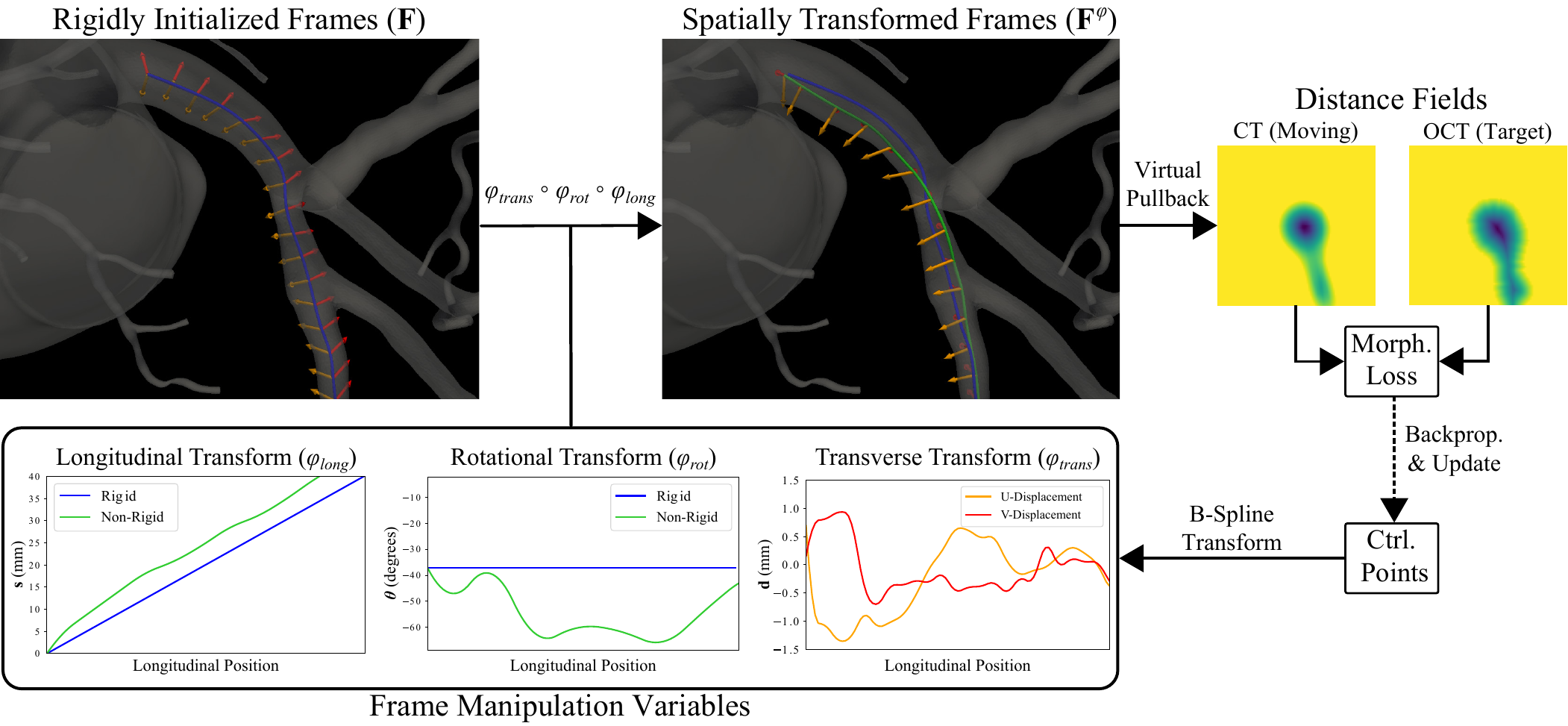}}
\caption{Overview of the spatial deformation acting on the {\color{black}rigidly initialized }virtual catheter path represented by a frame matrix \textbf{F}. The frames are acted upon by the longitudinal transform $\varphi_\text{long}$ that stretches and compresses the space between adjacent frames according to the longitudinal position vector $\mathbf{s}$. The rotational transform  $\varphi_\text{rot}$ rotates each orientation vector (red and orange arrows) about the catheter axis (blue line) according to the rotational vector $\boldsymbol{\theta}$. The transverse transform  $\varphi_\text{trans}$ shifts the frame centers in the direction of the rotated orientation vectors according to transverse displacement vectors $\mathbf{d}^u$ and $\mathbf{d}^v$. The spatially transformed frame matrix $\mathbf{F}^\varphi$ is used to sample cross sections from the CCTA lumen morphology that are compared to the target morphology derived from the OCT lumen. The parameters controlling the spatial transforms are then updated with gradient descent.}
\label{figur_Virtual}
\end{figure*}

{\color{black}Instead of directly manipulating the 3D positions and orientations of the frames $\mathbf{F}$ to produce $\mathbf{F}^\varphi$, each spatial transform takes as input one or more frame manipulation vectors that represent the stretching, twisting, and bending of the original virtual catheter path. As such, we define four frame manipulation vectors ($\mathbf{s},\boldsymbol{\theta},\mathbf{d}^u, \mathbf{d}^v$) representing 1) the arclength positions along the virtual catheter path $\mathbf{s}$, 2) the rotation angles of each frame $\boldsymbol{\theta}$ about the catheter, and 3) the in-plane transverse displacements $\mathbf{d}^u$ and $\mathbf{d}^v$ (see Fig. \ref{figur_Virtual}). This parametrization enables us to regularize the virtual catheter path to be smooth along the pullback, with independent smoothness constraints for each deformation type. To enforce such constraints, we control the frame manipulation vectors through B-spline deformations \cite{rueckert1999FFD} parametrized by a sparse set of control points.}

\begin{algorithm}
\begin{algorithmic}[1]\setstretch{1.2}
\caption{Non-rigid Co-registration}
\color{black}
\label{alg:nonrig_coreg}
\small
\Require $\mathbf{C},\vartheta, \mathbf{F}^\text{ori}$ \Comment{\text{Crop Indices}, \text{Rotation Angle}, Frames}
\Require $\mathbf{L}_{\text{CT}}^\text{3D}, \mathbf{L}_{\text{OCT}}^\text{pull}$ \Comment{Luminal Signed Distance Fields}
\State $\mathbf{s}_\text{init},\boldsymbol{\theta}_\text{init},\mathbf{d}^u_\text{init},\mathbf{d}^v_\text{init} \leftarrow \text{InitFrameVars()}$
\State $\mathbf{p}^s, \mathbf{p}^\theta, \mathbf{p}^{u}, \mathbf{p}^{v} \leftarrow \text{InitCtrlPts()}$
\State $\mathbf{x}^s, \mathbf{x}^\theta \leftarrow \text{InitRelVecs()}$ \Comment{Stretch \& Twist Vectors}
% \Algphase{Rigid Initialization}
\State $\mathbf{F},\mathbf{L}_{\text{OCT}}^\text{pull} \leftarrow \text{RigidInit}(\mathbf{C},\vartheta, \mathbf{F}^\text{ori} ,\mathbf{L}_{\text{OCT}}^\text{pull})$ %\Comment{Crop \& Rotate Frames}
% \Algphase{Optimization Loop}
\For{$i \in \{1...,$\text{Epochs}$\}$}\Comment{Optimization Loop}
\Algphase{Stretch Frames (Sec. \ref{nonrigid_methods_long})}
% \For{$j \in \{1...,m_s-1\}$} \Comment{Deform Control Points}
% \State  $p^s_j+=x^s_j+\sum^{j-1}_{k=0}x^s_k.$ 
% \EndFor
% \State $\textbf{s} \leftarrow \mathbf{B}^s\mathbf{p}^s$ \Comment{B-spline Deformation}
\State $\mathbf{p}^s \leftarrow \text{DeformCtrlPts(}\mathbf{x}^s\text{)}$\Comment{Equation \ref{eq:deform_long}}
\State $\textbf{s} \leftarrow \text{BsplineDeform(}\mathbf{s}_\text{init},\mathbf{p}^s\text{)}$ \Comment{Equation \ref{eq:matrixbasis}}
\State $\mathbf{F}^{s} \leftarrow \varphi_\text{long}(\mathbf{s})\circ \textbf{F}.$ \Comment{Equation \ref{eq:stretch}}

\Algphase{Twist Frames (Sec. \ref{nonrigid_methods_rot})}
% \For{$j \in \{1...,m_\theta-1\}$} \Comment{Deform Control Points}
% \State $p^\theta_j+=x^\theta_j+\sum^{j-1}_{k=0}x^\theta_k$
% \EndFor
% \State $\boldsymbol{\theta} \leftarrow \mathbf{B}^\theta \mathbf{p}^\theta$ \Comment{B-spline Deformation}
\State $\mathbf{p}^\theta \leftarrow \text{DeformCtrlPts(}\mathbf{x}^\theta\text{)}$\Comment{Equation \ref{eq:deform_theta}}
\State $\boldsymbol{\theta} \leftarrow \text{BsplineDeform(}\boldsymbol{\theta}_\text{init},\mathbf{p}^\theta\text{)}$ \Comment{Equation \ref{eq:matrixbasis_theta}}
\State $\mathbf{F}^{\theta} \leftarrow \varphi_\text{rot}(\boldsymbol{\theta})\circ \mathbf{F}^s$ \Comment{Equation \ref{eq:twist}}

\Algphase{Bend Frames (Sec. \ref{nonrigid_methods_trans})}
\State $\mathbf{d}^u  \leftarrow \text{BsplineDeform(}\mathbf{d}^u_\text{init},\mathbf{p}^{u}\text{)}$ \Comment{Equation \ref{eq:matrixbasis_trans}}
\State $\mathbf{d}^v  \leftarrow \text{BsplineDeform(}\mathbf{d}^v_\text{init},\mathbf{p}^{v}\text{)}$ \Comment{Equation \ref{eq:matrixbasis_trans}}
% \State $\mathbf{d^u} \leftarrow \mathbf{B}^{d,u} \mathbf{p}^{d,u}$ \Comment{B-spline Deformation}
% \State $\mathbf{d^v} \leftarrow \mathbf{B}^{d,u} \mathbf{p}^{d,v}$ \Comment{B-spline Defomation}

\State $\mathbf{F}^{\varphi} \leftarrow \varphi_\text{trans}(\mathbf{d}^u,\mathbf{d}^v)\circ \mathbf{F}^\theta$ \Comment{Equation \ref{eq:bend}}

\Algphase{Update Parameters}
\State $\mathbf{L}_{\text{CT}}^\text{pull} \leftarrow \text{VirtualCatheter}(\mathbf{F}^{\varphi}, \mathbf{L}_{\text{CT}}^{\text{3D}})$ 
\State $\mathcal{L} \leftarrow MSE\bigl(\text{clamp}(\mathbf{L}_{\text{CT}}^\text{pull}),\text{clamp}(\mathbf{L}_{\text{OCT}}^\text{pull})\bigl)$ \Comment{Loss}
\State $\mathbf{x}^s$, $\mathbf{x}^\theta, \mathbf{p}^{u},\mathbf{p}^{v} \leftarrow \text{Adam}(\nabla\mathcal{L})$ \Comment{Backprop \& Step}
\EndFor
\State \Return  $\mathbf{F}^{\varphi}$
\end{algorithmic}
\end{algorithm}

\subsubsection{Non-rigid longitudinal registration}
\label{nonrigid_methods_long}
The spatial transform $\varphi_\text{long}$ governing the inter-frame spacing along the virtual catheter takes in the arclength vector $\mathbf{s}$ and resamples a spline based on the initial centerline points $\mathbf{R}$ to produce an updated set of centerline coordinates $\mathbf{R}^s$. The frame poses ($\mathbf{T}^s$,$\mathbf{U}^s$, and $\mathbf{V}^s$) are then recalculated and used to update the frame matrix:
\begin{equation}
    \label{eq:stretch}
    \mathbf{F}^{s}=\varphi_\text{long}(\mathbf{s})\circ \textbf{F}.
\end{equation}
The initial arclength vector $\mathbf{s}_\text{init} \in\mathbb{R}^{n}$ is set to be monotonically increasing from 0 to 1, and is updated by a B-spline transform:
\begin{equation}
    \textbf{s}=\mathbf{B}^s\mathbf{p}^s,
    \label{eq:matrixbasis}
\end{equation}
in which $\textbf{s}\in\mathbb{R}^{n},\mathbf{B}^s\in\mathbb{R}^{n\times m_s},\mathbf{p}^s\in\mathbb{R}^{m_s}$ where $n$ is the number of frames and $m_s$ is the number of longitudinal control points. $\mathbf{B}^s$ is the univariate B-spline tensor and is pre-computed from the initial arclength vector $\mathbf{s}_\text{init}$, while $\mathbf{p}^s$ is the deformed control point vector that is initialized as a monotonically increasing vector of length $m_s$.

To account for the cumulative effect of catheter motor speed variation, we do not directly optimize the control points $\mathbf{p}^s$. Instead, we optimize for the relative stretch vector $\mathbf{x}^s \in \mathbb{R}^{m_s-1}$ that determines the cumulative displacement of each control point $\Delta {p^s_i}$, with the most proximal control point remaining fixed,
\begin{equation}
    \label{eq:deform_long}
    \Delta p^s_j=x^s_j+\sum^{j-1}_{k=0}x^s_k.
\end{equation}
To regularize the virtual catheter motion and prevent backward movement, the relative deformation of each control point $\Delta p^s_j$ is limited to a fraction of the distance between the control points through clamping. 
%Once the longitudinal control points $\mathbf{p}^s$ are deformed into a new configuration, the new arclength vector $\mathbf{s}$ is calculated through Equation \ref{eq:matrixbasis} and  used to resample the spline function to produce $\mathbf{R}^s$, upon which the frame vector matrices ($\mathbf{T}^s$,$\mathbf{U}^s$, and $\mathbf{V}^s$) are recalculated.

\subsubsection{Non-rigid rotational registration}
\label{nonrigid_methods_rot}
The rotational transform {\color{black}$\varphi_\text{rot}$ is applied to the longitudinally adjusted frames $\mathbf{F}^s$ and takes in the rotation angles $\boldsymbol{\theta}$ to produce the rotationally adjusted frames $\mathbf{F}^\theta$. This is done by rotating the longitudinally adjusted orientation vectors $\mathbf{U}^s$ and $\mathbf{V}^s$ about the tangent vector set $\mathbf{T}^s$.} 
\begin{equation}
    \label{eq:twist}
    \mathbf{F}^{\theta}=\varphi_\text{rot}(\boldsymbol{\theta})\circ \mathbf{F}^s
\end{equation}
The initial rotation vector $\boldsymbol{\theta}_\text{init} \in\mathbb{R}^{n}$ is initialized with zeros and is updated by a B-spline transform:
\begin{equation}
    \label{eq:matrixbasis_theta}
    \boldsymbol{\theta}=\mathbf{B}^\theta \mathbf{p}^\theta,
\end{equation}
where $\boldsymbol{\theta}\in\mathbb{R}^{n},\mathbf{B}^\theta\in\mathbb{R}^{n\times m_\theta}, \mathbf{p}^\theta \in\mathbb{R}^{m_\theta}$, where $m_\theta$ is the number of control points and $\mathbf{B}^\theta$ is a B-spline tensor that is pre-computed from the initial rotation vector $\boldsymbol{\theta}_\text{init}$. The rotational control point vector $\mathbf{p}^\theta$ is initialized as a zero vector and is updated similarly to the longitudinal control points, where the rotation defined for each control point is updated by a relative twist vector $\mathbf{x}^\theta \in \mathbb{R}^{m_\theta-1}$. The cumulative rotation value for each control point is therefore defined by:
\begin{equation}
    \label{eq:deform_theta}
    \Delta p^\theta_j=x^\theta_j+\sum^{j-1}_{k=0}x^\theta_k,
\end{equation}

\subsubsection{Non-rigid transverse registration}
\label{nonrigid_methods_trans}
{\color{black}The transverse transform $\varphi_\text{trans}$ is applied to the rotationally adjusted frames  $\mathbf{F}^\theta$ and takes as input displacement magnitude vectors $\mathbf{d}^u$ and $\mathbf{d}^v$ to produce the final frames $\mathbf{F}^\varphi$. This is done by displacing the rotationally aligned centerline points $\mathbf{R}^\theta$ along $\mathbf{U}^\theta$ and  $\mathbf{V}^\theta$ to obtain $\mathbf{R}^\varphi$.}

\begin{equation}
    \label{eq:bend}\mathbf{F}^{\varphi}=\varphi_\text{trans}(\mathbf{d}^u,\mathbf{d}^v)\circ \mathbf{F}^\theta
\end{equation}

In contrast to the longitudinal and rotational transforms, the transverse transform $\varphi_\text{trans}$ consists of two separate operations ($\varphi_\text{trans}^u$ and $\varphi_\text{trans}^v$) that control the transverse displacement of the catheter path away from the artery center in orthogonal directions.
\begin{equation}
    \varphi_\text{trans}(\mathbf{d}^u,\mathbf{d}^v)=\varphi_\text{trans}^u(\mathbf{d}^u)\circ\varphi_\text{trans}^v(\mathbf{d}^v),
\end{equation}

The initial in-plane transverse displacements $\mathbf{d}^u$ and $\mathbf{d}^v$ are initialized to be zero and are calculated by the following relation:
\begin{equation}
    \label{eq:matrixbasis_trans}
    \mathbf{d^u}=\mathbf{B}^{u} \mathbf{p}^{u}.
    \quad \text {and}\quad 
    \mathbf{d^v}=\mathbf{B}^{u} \mathbf{p}^{v}.
\end{equation}
where each displacement vector is controlled by control points $\mathbf{p}^{u}$ and $\mathbf{p}^{v}$. The virtual catheter is initialized to stay close to the centerline by setting the two control point vectors as zero-vectors of length $m_d$ each. In contrast to longitudinal and rotational registration, we directly optimize the control points as the artery wall constrains the cumulative transverse displacement of the catheter.

% \newpage

\subsection{Evaluation}
\label{sec:evaluation}
\subsubsection{Image data}
\label{dataset}
To evaluate our proposed co-registration framework, a dataset consisting of 40 matched OCT and CCTA image pairs from five different clinical centers were selected, all originating from the Precise Percutaneous Coronary Intervention Plan (P3) study \cite{nagumo2021rationale}. As each OCT pullback image consisted of 375 frames, the intravascular imaging dataset comprised of approximately 15,000 image frames before excluding frames with poor image quality. The OCT lumen in every frame was manually annotated by trained cardiologists, and continuous segments of the OCT pullback with poor lumen segmentations due to residual blood or catheter housing were manually excluded. Further,{\color{black} as no manual annotations were available}, the vessel wall borders in every OCT frame were segmented using a convolutional neural network through a U-net architecture \cite{ronneberger2015unet}. Details of the network, training, and validation can be found in section \ref{unetsection}. The lumen and vessel wall segmentations were re-sampled to represent a 3D image of dimensions $(96 \times 96 \times n)$ with an in-frame resolution of 80 micrometers and an out-of-frame resolution of 0.4 mm (sampling every other longitudinal frame). {\color{black}The segmentations of the lumen and vessel wall were used to produce the SDFs $\mathbf{L}_\text{OCT}^\text{pull}$ and $  \mathbf{W}_\text{OCT}^\text{pull}$ using the fast marching method \cite{treister2016fastmarching}}. All utilized intravascular pullback sections  were manually deemed to sufficiently visualize the artery. For the CCTA data, a 3D {\color{black} surface mesh of the coronary tree for each patient was provided by a previously validated virtual planner \cite{sonck2022clinical}. These meshes were produced by a deep learning algorithm and are minimally corrected through human annotators}. The 3D models are used to produce high-resolution SDFs of the lumen and vessel wall  $\mathbf{L}_\text{CT}^\text{3D} $ and $ \mathbf{W}_\text{CT}^\text{3D}$ with a resolution of 0.25 mm along each axis and a shape of of $(768 \times 768 \times 482)$. {\color{black} The CCTA SDFs were stored as truncated signed distance fields, only containing positive distance values up to 2mm. This was done to enable high-fidelity sampling through virtual catheterization while also reducing the pre-processing cost. The vessel centerlines were semi-automatically obtained by annotating the start and end points of each artery and using them as input to VMTK \cite{antiga2008imagevmtk}}.

\subsubsection{Co-registration evaluation}
\label{sec:landmarks}
In order to evaluate the performance of the non-rigid registration, 114 bifurcations were manually marked {\color{black}by human experts} in the OCT pullback as well as in the rigid and non-rigid virtual pullback segmentations generated from the CCTA data. Bifurcations were defined as the last image frame before a coronary artery splitting into two branches could be seen. The landmark annotations were first annotated before non-rigid registration for the rigidly aligned data belonging to both modalities. Specifically, bifurcations that were common to both modalities had their frame numbers recorded for validation of the non-rigid registration algorithm. The initially annotated bifurcations in the CCTA pullback were then re-annotated after non-rigid registration. Longitudinal validation was conducted by comparing the frame number of a bifurcation in the OCT data with the equivalent bifurcation frame number in the virtual pullback before and after non-rigid registration. 

In order to validate the non-rigid rotational registration, the manually annotated bifurcation angles for the OCT pullback and the virtual pullback were compared before and after rotational registration. As the bifurcation angle between bifurcation sections that were not longitudinally matched is expected to be uncorrelated, only bifurcations that had a frame mismatch below a certain number of frames were considered for qualatative analysis of rotational accuracy. The longitudinal mismatch threshold was chosen as double the kernel size of the Gaussian filter applied to the SDF (six frames).

\begin{figure*}[!ht]
\centerline{\includegraphics[width=0.8\textwidth]{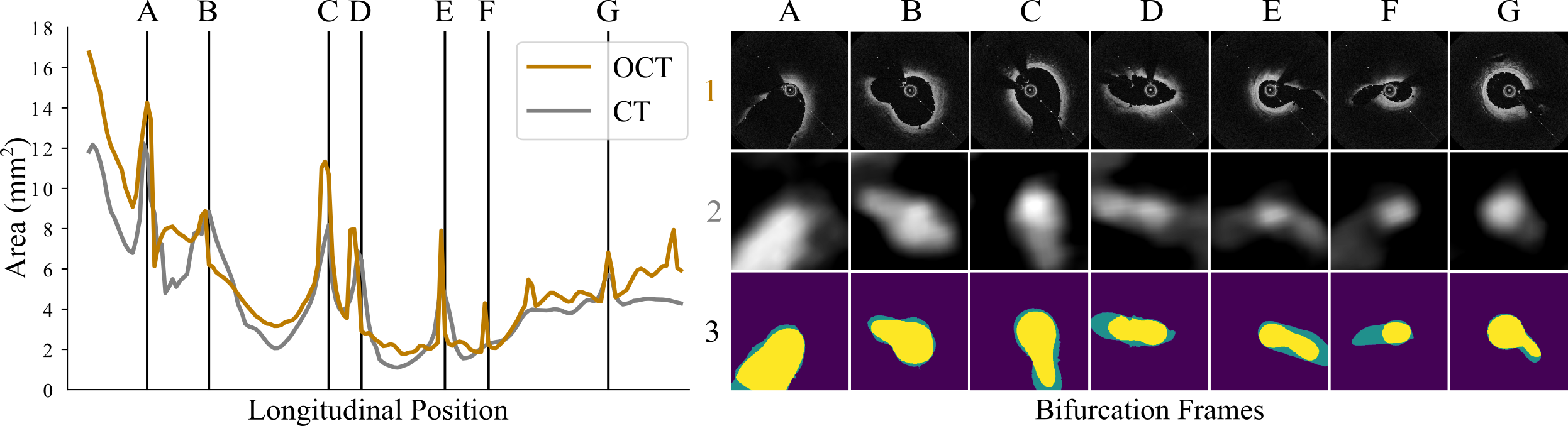}}
\caption{Qualitative results for a single co-registered case. The left plot displays the area along the artery for the non-rigidly registered CCTA (gray) and the OCT images (gold). The right plot displays the bifurcation zones (Sections A-G) that are marked and labeled for further visualization. Bifurcation frames from the CT, OCT, and overlapped segmentation maps are presented in the first, second, and third row for qualitative comparison.}
\label{figure_Main}
\end{figure*}

\subsubsection{Implementation details}
\label{sec:implentation}
The rigid longitudinal registration parameters were kept the same as the previous study \cite{karmakar2020detailed}. For the rigid rotational registration, the {\color{black}number of circumferential rays for each frame $\gamma$ was set to 30}. For the non-rigid registration, the gradient descent-based optimization procedure was implemented in PyTorch with the Adam optimizer \cite{kingma2014adam} with the default hyper-parameters. {\color{black} The parameters optimized were the relative stretch vector $\mathbf{x}^s$, the relative twist vector $\mathbf{x}^\theta$, and the control points associated with the transverse displacements $\mathbf{p}^{d,u}$, and $\mathbf{p}^{d,v}$. A learning rate of 0.001 was used for the non-rigid longitudinal parameters  while the non-rigid rotational and non-rigid transverse parameters had a learning rate of 0.01}. This was done to encourage rough longitudinal alignment of bifurcations early in the optimization process. Each co-registration procedure was run for a minimum of 200 iterations to ensure convergence. The number of control points $m_s$, $m_\theta$, and $m_d$ were chosen to be 30, 20, and 60 respectively, to match the frequency of variation for each aspect of catheter motion. The relative deformation of the longitudinal control points $\mathbf{p}_s$ was limited to be 0.35 times the inter-point distance. {\color{black} Finally, the Gaussian kernel used to smooth the SDFs was implemented with a standard deviation of 0.1 and a kernel size of three voxels.}

\begin{figure}[!h]
   \begin{minipage}{0.45\textwidth}
     \centering
     \includegraphics[width=.7\linewidth]{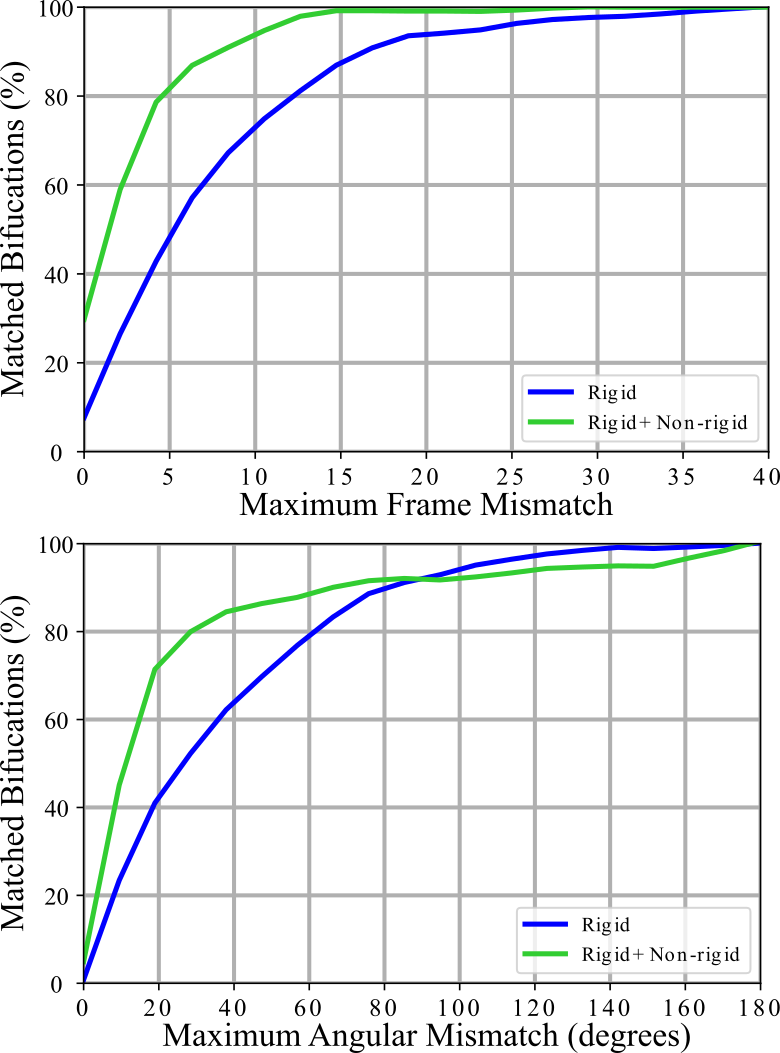}
     \caption{Quantitative results comparing rigid and non-rigid co-registration in longitudinal and rotational directions with varying degrees of misalignment. Mismatch plots exhibit the \% of matched bifurcations with increasing longitudinal (top) and rotational (bottom) alignment mismatch criteria (x-axis).}\label{figuremismatch}
   \end{minipage}\hfill

%\begin{comment}
   \begin{minipage}{0.45\textwidth}
     \centering
     \includegraphics[width=.7\linewidth]{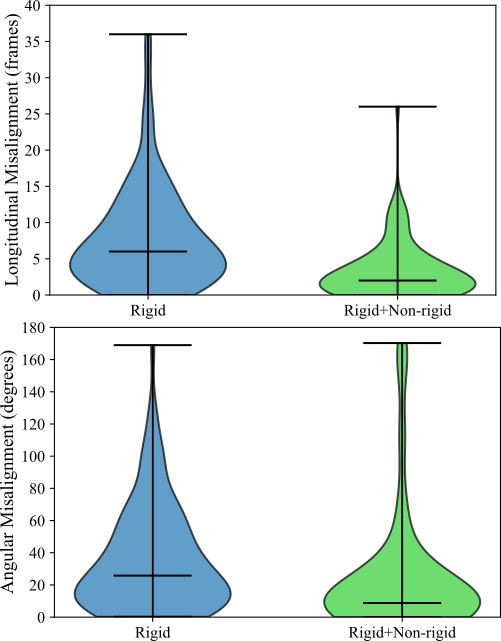}
    \caption{Violin plots comparing rigid and non-rigid co-registration in longitudinal and rotational directions. horizontal bars mark median and extremes. First row compares longitudinal bifurcation frame mismatch before  and after non-rigid registration. Second row compares bifurcation angular mismatch before and after non-rigid registration. Bifurcations that were longitudinally matched within six OCT frames after non-rigid registration were plotted in the second row.}
    \label{figureviolin}
   \end{minipage}
\end{figure}

\subsubsection{Baseline approach}
\label{sec:baseline}
The most commonly used automatic co-registration methodologies employed for coronary artery registration have been discrete optimization approaches such as DTW and DP. In order to evaluate the performance of our longitudinal and rotational co-registration framework against state-of-the-art discrete approaches, we applied the methodology described in our previous work by Karmakar et. al \cite{karmakar2022framework} on the same dataset. The approach utilizes DTW to longitudinally align two coronary imaging modalities and DP to rotationally align each frame. We utilized a window length of four frames and recorded identical alignment metrics for 114 matched bifurcations in the dataset. The non-rigid registration algorithm was applied after our rigid longitudinal registration step described in section \ref{Rigid Registration}. 
\subsubsection{Vessel wall segmentation model}\label{unetsection}
In this study, our rigid rotational registration procedure required approximate vessel wall segmentations. As the rotational registration initialization was only required to be approximate, the segmentations were not required to be high-fidelity or topologically accurate. Therefore, a segmentation network was trained to produce vessel wall label maps from 2D intravascular OCT frames. We utilized a U-net architecture with a resnet50 encoder \cite{he2015deep}. Our dataset consisted of a mixture between a previously annotated dataset \cite{olender2018mechanical} consisting of 8 OCT pullbacks and two newly annotated OCT pullbacks from the P3 trial dataset, totaling 1500 2D OCT frames. 105 frames corresponding to one entire pullback were held out for validation. For augmentation, we utilized random affine transformations with a rotational range of [0, 180] degrees and a scale range of [0.6, 1.4]. A learning rate of 0.0001 was used in tandem with the Adam optimizer. When applied to the validation set, the model exhibited a precision of 0.85 and a binary dice score of 0.78. 

% \newpage
\section{Results}
\label{sec:results}
\begin{figure*}[!h]
\centerline{\includegraphics[width=0.8\textwidth]{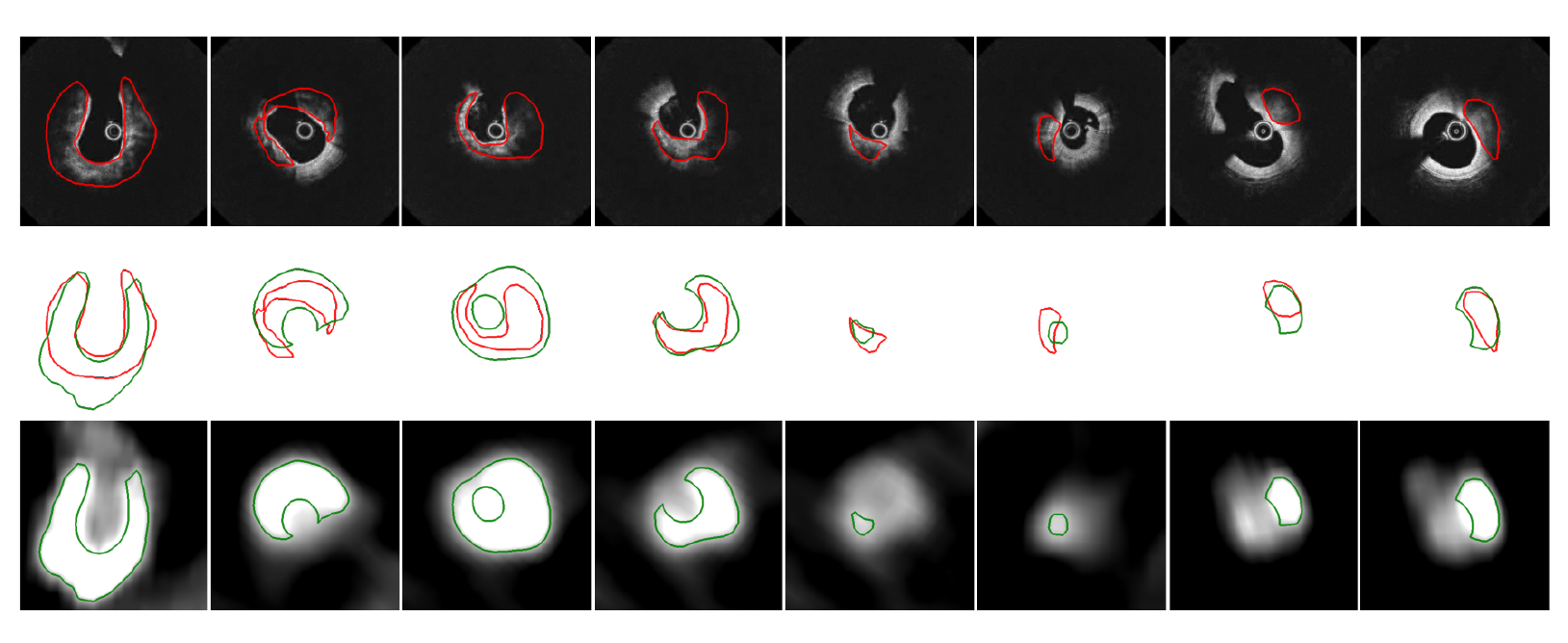}}
\caption{Qualitative results comparing calcium annotations between CCTA (first row, obtained by thresholding) and OCT (third row, obtained by manual annotation) for selected frames with sufficient luminal alignment. Middle row shows superimposed calcium annotations for OCT (red) and CCTA (green).}
\label{figure_calc}
\end{figure*}

\subsection{Longitudinal registration}
Longitudinal registration can be qualitatively seen in Fig. \ref{figure_Main}, where the non-rigid registration process aligned the majority of common bifurcations in both imaging modalities. The improvements over rigid registration are further visualized by a longitudinal mismatch plot (Fig. \ref{figuremismatch}A), revealing that after rigid alignment, the percentage of bifurcations matched within two, four, and six frames are 26.3, 42.1, and 57.9\%, respectively, while after non-rigid alignment, these values increase to 60.5, 78.9, and 86.8\%. Examining the mismatch distribution through the longitudinal mismatch violin plot in Fig. \ref{figureviolin}, it can be shown that using rigid registration alone, there exists a significant variability in longitudinal mismatch, with the median mismatch being six frames. However, after non-rigid alignment (Fig. \ref{figureviolin}), distinct improvement can be observed with a majority of bifurcations experiencing a decrease in longitudinal mismatch, with the median mismatch decreasing to 2 frames.  Table \ref{table_results_comb} further demonstrates the effect of non-rigid registration, in which the mean frame difference after rigid registration was 7.9 frames {\color{black}(1.58 mm)} and subsequently decreased to 3.3 frames {\color{black}(0.66 mm)} after non-rigid registration. {\color{black} Statistical significance between the longitudinal non-rigid and rigid registration error was determined by a  Wilcoxon signed rank test ($p$ \( < 0.001 \))}.

\subsection{Rotational registration}
Examination of the individual bifurcating frames in Fig. \ref{figure_Main} for the CCTA (row 1) and OCT (row 2) frames indicates qualitative rotational and transverse alignment between both imaging modalities as evident from the raw images and the overlapped segmentations (row 3). Furthermore, Fig. \ref{figure_calc} demonstrates co-registration of calcific inclusions in regions adjacent to properly aligned bifurcations. Rotational registration plots in Fig. \ref{figureviolin} quantitatively demonstrate that many bifurcations exhibit high levels of angular misalignment, with a median misalignment of 25.8 degrees. After non-rigid alignment, a significant number of misaligned bifurcations were enhanced in terms of their alignment, bringing the median mismatch down to 8.8 degrees. Examination of the rotational mismatch plot (Fig. \ref{figuremismatch}) quantitatively demonstrates an increase in the percentage of bifurcations aligned up to an angular mismatch of 10, 20, and 30 degrees from \% values of 25.3, 40.4, and 52.3 to 51.5, 69.7, and 79.8\%, respectively. The mean value of the angular mismatch before and after non-rigid alignment is reported in Table \ref{table_results_comb}, in which the mean angular mismatch decreases from 36.0 to 28.6 degrees. {\color{black} Statistical significance between the rotational non-rigid and rigid registration error was determined by a  Wilcoxon signed rank test ($p$ \( < 0.001 \))}.

\subsection{Comparison with baseline}
A direct comparison of our virtual catheter method with state-of-the-art discrete optimization approaches can be seen in Table \ref{table_results_comb}.
Comparing the virtual catheter method to a discrete optimization approach for longitudinal registration, it can be seen that DTW produces significantly poorer results in longitudinal registration, with the longitudinal mismatch of 11.7 frames (2.34 mm) being higher than rigid longitudinal registration average of 7.9 frames. Comparing the virtual catheter method to using DP for rotational registration, discrete optimization algorithms exhibit poor performance for CT-OCT rotational registration (angular mismatch of 77.9 degrees) which is higher than the angular mismatch after rigid rotational registration alone. {\color{black} Statistical significance between registration errors was determined by a  Wilcoxon signed rank test ($p$ \( < 0.001 \))}.

\begin{table}[ht!]
%\label{Coregistration_Table}
  %\begin{center}
    \caption{Accuracy of co-registration approaches applied to CT-OCT image registration. Average errors and standard deviations in longitudinal (frames) and rotational (degree) directions. All approaches in CT-OCT are evaluated on the same dataset}
    \label{table_results_comb}
      \resizebox{\columnwidth}{!}{\begin{tabular}{c|c|c|c|c} % <-- Alignments: 1st column left, 2nd middle and 3rd right, with vertical lines in between
      \textbf{Method} & \textbf{Modalities} & \textbf{Subjects} & \textbf{\shortstack{Frame \\ mismatch}} &\textbf{\shortstack{Degree \\ mismatch}}\\
      \hline
      \cite{karmakar2022framework} & OCT-OCT & 9  & $0.9 \pm 0.8$ & $7.7 \pm 6.7$\\
      \cite{tsiknakis2023octregistration} & OCT-OCT & 21  & $5.6 \pm 6.7$ & $1.2 \pm 0.81$\\
      \hline
      \cite{karmakar2022framework} & OCT-IVUS & 7  &  $1.45 \pm 0.7$ & $29.1 \pm 23.2$\\
      \cite{molony2016evaluation} & OCT-IVUS & 12  & $ 5.0 \pm 6.2$ & $17.8 \pm 21.9 $\\
      \hline  
      \cite{karmakar2022framework} & CT-OCT & 40  & $ 11.7 \pm 12.1 $ & $77.9 \pm 61.0$ \\
      Ours (Rigid) & CT-OCT & 40  &  $7.9 \pm 7.1 $ & $36.0 \pm 31.9 $\\
      Ours (Non-rigid) & CT-OCT & 40 &  \ $3.3 \pm 3.9 $ & \ $28.6 \pm 40.9$\\
    \end{tabular}}
  %\end{center}
\end{table}

\section{Discussion}
The aim of the current study was to develop a semi-automatic registration algorithm to align CCTA and intravascular images given the equivalent vessel morphology and a CCTA centerline as guiding inputs. Specifically, we propose a novel registration process that involves finding the optimal rigid and non-rigid spatial transforms applied to a virtual catheter moving through the CCTA image, aligning both modalities. Our results indicate that our co-registration methodology can align CCTA and OCT frames with a high degree of fidelity, as evidenced by the alignment of reference landmark annotations (Fig. \ref{figure_Main}). Further, our results underline the critical importance of a non-rigid registration step, with significant enhancement in both  longitudinal and rotational alignments as seen when comparing rigid vs. non-rigid alignments in Table \ref{table_results_comb}. We demonstrate that for the majority of bifurcations, our framework is able to improve the longitudinal and rotational alignment of common bifurcations within the CCTA and OCT images (Fig. \ref{figureviolin}). Lastly, we demonstrate the added value of our approach as compared to state-of-the-art alternatives, with a head-to-head comparison to previously developed discrete optimization alignment algorithms (Table \ref{table_results_comb}). This comparison demonstrates that discrete optimization approaches for longitudinal and rotational alignment suffer a significant drop in alignment quality when applied for the task of CT-OCT co-registration. Meanwhile, our approach maintains performance metrics in line with intravascular-intravascular image registration.
%Comparison to other approaches
\subsection{Related work}
\label{sec:related}
Currently, a majority of CCTA studies that validate their findings with intravascular images have used manual registration based on fiduciary landmarks such as bifurcations or large calcifications \cite{carlier2014new,tu2011fusion,hebsgaard2015co,uzu2019lumen}. In comparison, our approach implicitly matches nearby bifurcations using morphological representations of the CCTA and OCT lumen. Other approaches that automatically register intravascular-to-intravascular modalities have in the past relied on DTW \cite{molony2016evaluation,karmakar2022framework}, to maximize longitudinal and rotational alignment of separate intravascular pullbacks.

Direct numerical comparison of reported co-registration accuracy across published approaches is inherently difficult as co-registration accuracy is highly dependent on the specific datasets used as well as which modalities are being co-registered. For example, co-registration accuracy is higher for single-modality datasets (OCT-OCT) compared to datasets that include multiple modalities (OCT-IVUS) (Table \ref{table_results_comb}). {\color{black} Moreover, this problem is made more difficult as many co-registration studies are conducted on small and private datasets consisting of few patients. In contrast, we leverage a multi-center dataset of 40 patients that is significantly larger than the average dataset size of comparable prior studies. We also directly compare with our previously developed discrete optimization algorithm \cite{karmakar2022framework} to control for dataset variability, finding that our prior work produced significantly worse longitudinal and rotational alignment compared to the virtual catheter method for the case of CT-OCT registration (Table \ref{table_results_comb}). In contrast, our developed methodology achieves similar results to studies involving intravascular-intravascular registration (Table \ref{table_results_comb}).}

\subsection{Methodological adaptations}
\label{sec:adaptations}
\color{black} The task of co-registering CCTA and OCT images presents several unique difficulties for discrete registration algorithms. Our framework has several features that were designed to mitigate such challenges. First, the {\color{black}comparatively }low resolution of CCTA images induces a circular bias in the already circular lumen segmentations (see Fig. \ref{figure_Main}), as well as a tendency to miss small bifurcations. Such circularly symmetric regions create zones of longitudinal and rotational ambiguity along the pullback. Our approach minimzes this effect by formulating the longitudinal and rotational transforms in terms of regularized and smooth B spline deformations. As such, the optimization procedure is mainly guided by the alignment of prominent non-symmetric features such as bifurcations, rather than the circularly symmetric lumen segments. This incentivizes the rotational alignment of all non-bifurcating lumen frames that are in proximity to their matched bifurcations (Fig. \ref{figure_Main}).

Another significant issue faced in previous rotational co-registration algorithms \cite{karmakar2022framework,molony2016evaluation} is that lumen bifurcations are only able to contribute to rotational alignment if they exist within the same frame. As such, poor longitudinal alignment of bifurcations was a significant contributing factor to the poor performance of our previously developed DP algorithm for rotational co-registration (Table \ref{table_results_comb}). Our current framework, in contrast, minimizes this dependency through the use of a 1D Gaussian smoothing kernel applied longitudinally over the OCT morphology. Longitudinal smoothing allows single-frame bifurcations to appear in adjacent frames and smooths the loss surface such that bifurcations in the different modalities can be better aligned (Fig. \ref{figure_Main}.

%Use of transverse transform
Lastly, many co-registration methods normalize the position of the lumen by the artery centroid \cite{uzu2019lumen,karmakar2020detailed,karmakar2022framework,molony2016evaluation}. While such an approach manages to align CCTA and OCT frames with circularly symmetric lumens, it fails to align equivalent frames with bifurcations, due to differing centroids between the modalities. Moreover, centering the image around the lumen centroids can cause the algorithm to mistakenly align bifurcations 180-degrees from the correct orientation. In our current framework, we instead choose to jointly optimize for the transverse displacements of the catheter path frames in addition to the longitudinal and rotational displacements, which allows for the bifurcations in both modalities to be anchored around the OCT catheter location and enables near pixelwise alignment of the lumen (Fig. \ref{figure_Main}) and plaque constituents such as calcium (Fig. \ref{figure_calc}).

{\color{black} In contrast to approaches that minimize image similarity for co-registration \cite{viola1997alignmentsandyMI}, our morphology-based approach is agnostic to the specific intravascular imaging modality provided that luminal segmentations are available. As such, it is likely that our non-rigid algorithm can be readily extended to co-register CCTA and IVUS images, as IVUS can visualize the lumen with similar quality compared to OCT images, albeit with a minor bias towards over-estimating the lumen area \cite{kubo2013octsizeivus,huang2022calcifiedoctivus}. However, the applicability of rigid rotational registration with IVUS-derived vessel wall segmentations is an open question. On one hand, calcified plaque can significantly effect the visualization of the wall through acoustic shadowing \cite{huang2022calcifiedoctivus} which can impact the rotational registration accuracy after rigid registration. On the other hand, rigid rotational registration must only produce an adequate initialization for the non-rigid registration algorithm, which may be insensitive to non-extensive imaging artifacts. Moreover, rigid registration can be approximated with the annotation of a single fiduciary landmark, meaning that our non-rigid algorithm can nonetheless accelerate registration without relying on IVUS-derived vessel wall segmentations.}
\subsection{Limitations}
\label{sec:limitations}
\color{black}
%Limitations
Though very promising for clinical applications, our developed approach has a number of limitations. First, the non-rigid spatial transform acting on the virtual catheter is found through gradient-based optimization, requiring that the rigid initialization brings landmarks sufficiently close such that proper matching is ensured. For example, common bifurcations that have a frame mismatch of more than six frames (corresponding to the longitudinal smoothing kernel) are expected to be uncorrelated in terms of orientation. This issue can be mitigated by training a neural network to predict the spatial transform needed to align the two modalities. As our developed spatial transforms are differentiable, they can be integrated into deep learning workflows with relative ease. Another limitation is the dependence of non-rigid registration on the lumen segmentations. The lumen estimation for bifurcations is expected to be accurate for both modalities and as such, ensures good registration accuracy for regions that include many such landmarks. However, due to the low resolution of CCTA as compared to intravascular modalities, the lumen estimation tends to be highly circular in vessel sections without bifurcations. Accordingly, it is expected that rotational co-registration certainty increases with bifurcation proximity but decreases in regions that contain highly circular luminal profiles. In the future, co-registration accuracy can likely be improved by including contextual information relating to the vessel wall such as lesion morphology as a supervisory signal in the loss function.Furthermore, the use of a pixel-wise loss as a surrogate for luminal alignment may not necessarily result in optimal alignment of lumen bifurcations. In the future, this issue can be mitigated by introducing an orientation loss to bias the spatial transform to rotationally align bifurcations. Lastly, regularizing the spatial transform and smoothing the SDFs can create difficulties in localizing landmarks up to frame-wise precision. This can be seen in the area curve in Fig. \ref{figure_Main} section B with the slightly mismatched bifurcation in the longitudinal direction. The localization capabilities of the algorithm can be improved by introducing multi-scale deformation steps where finer control point grids can be recursively used as the basis for the spatial transform.

\section{Conclusion}
\label{sec:conclusion}
{\color{black} We present a semi-automatic algorithm for the co-registration of CCTA and intravascular images. We formulate rigid and non-rigid registration algorithms to reconstruct the 3D path of the intravascular catheter, enabling a frame-to-frame comparison between modalities. Specifically, we use automatic differentiation to optimize for the virtual catheter path throughout the CCTA-derived lumen that recapitulates the lumen morphology as found in the intravascular image. Key to our approach is a differentiable spatial transform that models the non-rigid motion of the virtual catheter in the longitudinal, rotational, and transverse directions. Our non-rigid registration algorithm enables the creation of matched multi-modal datasets for various clinical applications and can be used in machine learning-based frameworks.

}

\bibliographystyle{IEEEtran}
\bibliography{bib}
\end{document}